\documentclass{article}
\usepackage[latin1]{inputenc}
\usepackage{graphics}
\usepackage{amsfonts, amssymb, latexsym, amsmath}
\def\author #1{\expandafter\def\expandafter\@aabuffer\expandafter
        {\@aabuffer\small\rm\uppercase{#1}\relax\par
        \vspace*{2pt}}}
\def\address#1{\expandafter\def\expandafter\@aabuffer\expandafter
        {\@aabuffer\small\it #1\relax\par
        \vspace*{10pt}}}
\begin{document}
\title{Recursive Weak- and Strong Coupling Expansions in a Cosine Potential.}
\author{B.~Hamprecht}
\address{Institut für Theoretische Physik, Freie Universität Berlin,\\
Arnimallee 14, D-14195 Berlin, Germany\\
{\scriptsize e-mails: bodo.hamprecht@physik.fu-berlin.de}}

\maketitle

\abstract{For the Cos(2x)-Potential the coefficients of the weak- and strong coupling perturbation series of the ground state energy are constructed recursively. They match the well-known expansion coefficients of the Mathieu equation's characteristic values. However presently there is no physically intuitive method to extract the coefficients of the strong coupling series from those of the weak one. The standard rule while giving exellent results for the anharmonic oscillator fails completely in this case.}

\section{Introduction}
The perturbative treatment of quantum-mechanical or field-theoretical
problems renders in general results in the form of divergent infinite
power series in some coupling constant $g$.  Various
resummation schemes have been applied to them, aiming at finite
results for all values of the coupling constant $g$. For the anharmonic oscillator exellent results are known.(See e.g.
Chap.~16 of the book of Kleinert and Frohlinde\cite{Verena} and the references therein).  Since the real value of the resumation schemes lies in their generalization and application to field theories, it is worth while to study, how generally applicable they are. But this requires models for which higher oder coefficients of the perturbation expansion are available. This is practically only the case, if the coefficients can be constructed recursively. Known models with this property are all closely related to the harmonic oscillator. They permit weak coupling expansions only. Here we present a model, which is based instead on the infinite square well potential. It permits the recursive construction of a weak- as well as a strong coupling series. Both series seem to have a finite radius of convergence and agree with numerical data. A mapping of one series to the other, based on physical intuition, is not known at present.

\section{The Model}
Consider the one dimensional Schrödinger equation:
\begin{equation}
\label{Schr}
-\frac{1}{2}\Psi''+(V_0+g\;V_1)\Psi=E(g)\;\Psi, 
\end{equation}
where the coupling constant $g$ may take on positive or negative values. Its unperturbed potential is chosen to be the infinite square well:
\begin{equation*}
\label{SqW0}
V_0( x ) = \left \{ \begin{array}{ll}
0 & \mbox{if}\quad |x|<\frac{\pi}{2}\\
\infty \quad & \mbox{otherwise}
\end{array}\right \}
\end{equation*}
As perturbing potential we take:
\begin{equation*}
\label{SqW1}
V_1( x ) = cos(2x)
\end{equation*}
The ground state energy $E_0(g)$ is known to be the characteristic value $b_1(g)$ of the Mathieu equation. We show now, how the expansion coefficients of $E_0(g)=b_1(g)$ can be obtained recursively in a quatum mechanical context.

\section{The Weak Coupling Series}
The recursion relation for the weak coupling coefficients $\epsilon_i$ in the series 
\begin{equation*}
\label{E0}
E_0(x) = \sum_{i=0}^{\infty} \epsilon_i\;g^i
\end{equation*}
for the ground state energy will be obtained by the method of Hamprecht and Pelster \cite{HaPe}. To start off, we need to know the matrix elements of the perturbing potential $V_1(x)$ in the basis of unperturbed eigenfunctions: 
\begin{equation*}
\label{EF}
\Psi_n(x) = \sqrt{\frac{2}{\pi}}\left \{ \begin{array}{ll}
cos\;n\,x & \mbox{for}\;n=1,3,5,\ldots\\
sin\;n\,x & \mbox{for}\;n=2,4,6,\ldots\\
\end{array}\right \}
\end{equation*}
They can be obtained by simple integration:
\begin{equation*}
\label{ME}
V_{n,m}:=\langle \Psi_n|cos(2x)|\Psi_m \rangle =\frac{1}{2} \left \{ \begin{array}{rl}
1 \; & \mbox{if}\quad m=n=1\\
1 \; & \mbox{if}\quad |n-m|=2 \; \mbox{and $n$ odd}\\
-1\; & \mbox{if}\quad |n-m|=2 \; \mbox{and $n$ even}\\
0\;  & \mbox{otherwise}\\
\end{array}\right \}
\end{equation*}
Since the matrix $V_{n,m}$ is band-diagonal, the method of Hamprecht and Pelster applies and produces the well known \cite{ASt} weak coupling coefficients as shown in {\it table~\ref{TabI}}. They are of Borell type in the sense, that tripletts of consecutive coefficients alternate in sign. Also they decrease with an average ratio of about $4$, so that the weak coupling series will have a radius of convergence of $|g|\simeq 4$. The recurrence relation for the $\epsilon_i$ can be found in appendix A.
\begin{table}[htp!]
\caption[TabI]{ The first few coefficients of the weak coupling expansion for the ground state of the $cos(2x)$-potential inside an infinite square well.}
\label{TabI}
\begin{center}
\begin{tabular}{||c|c||}\hline \hline
\multicolumn{2}{||c||}{\bf Square Well with $Cos(2x)$} \\ \hline
\multicolumn{2}{||c||}{\it Weak Coupling} \\ \hline
$i $ & $\epsilon_i$  \\ \hline \hline
$0 $ & $ 1/2 $ \\ \hline
$1 $ & $ 1/2 $ \\ \hline
$2 $ & $ -1/16 $ \\ \hline
$3 $ & $ -1/128 $ \\ \hline
$4 $ & $ -1/3\,072 $ \\ \hline
$5 $ & $ 11/73\,728 $ \\ \hline
$6 $ & $ 49/1\,179\,648 $ \\ \hline
$7 $ & $ 55/18\,874\,368 $ \\ \hline
$8 $ & $ -83/70\,778\,880 $ \\ \hline
$9 $ & $ -12\,121/30\,198\,988\,800 $ \\ \hline
$10 $ & $ -114\,299/3\,261\,490\,790\,400 $ \\ \hline \hline
$11 $ & $ 192\,151/15\,655\,155\,793\,920 $ \\ \hline \hline
$12 $ & $ 83\,513\,957/17\,533\,774\,489\,190\,400 $ \\ \hline \hline
\end{tabular}
\end{center}
\end{table}

\section{The Strong Coupling Series}
\begin{figure}
\caption[FigI]{The first few approximations of order $n=3,5,7,9$ to the rescaled ground state energy $E_0(g)/g$, which tends to the constant value $1$ for $g\rightarrow \infty$ are shown in comparison to numerical results (big dots). For small $|g|$ the weak coupling series (dotted lines) give good agreement and for large $|g|$ the strong coupling approximations (solid lines) fit very well.}
\begin{center}
\label{FigI}
\setlength{\unitlength}{1cm}
\begin{picture}(19, 7)
\scalebox{1.}[1.]{\includegraphics*{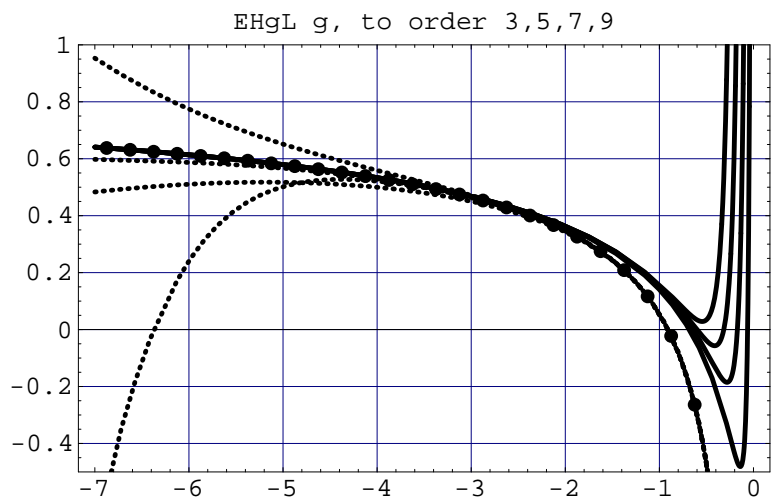}}
\end{picture}
\end{center}
\end{figure}

\begin{figure}
\caption[FigII]{The same data as in {\it figure~\ref{FigI}}, but here for the unscaled ground state energy $E_0(g)$ and for an extended intervall of the coupling constant $g$, including positive values as well.}
\begin{center}
\label{FigII}
\setlength{\unitlength}{1cm}
\begin{picture}(19, 7)
\scalebox{1.}[1.]{\includegraphics*{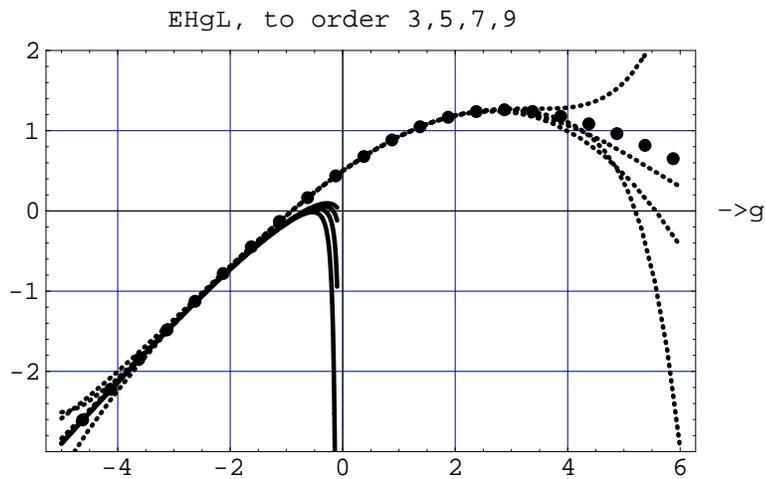}}
\end{picture}
\end{center}
\end{figure}

In this section we investigate the limit $g \rightarrow -\infty$. Scaling the x-axis with $x\rightarrow \alpha\,x$ with $\alpha=1/\sqrt{|g|}$ transforms the Schrödinger equation (\ref{Schr}) into:
\begin{equation}
\label{Schr2}
-\frac{1}{2}\Psi''+2\,sin(2\alpha \,x)\,\Psi=e\,\Psi, 
\end{equation}
where $E_0(g)=|g|(e-1)$. A solution to this eqaution will be found by an expansion of $Log(\Psi)$ in powers of $\alpha$. The square well barriers are withdrawn to $|x|=\pi/(2\sqrt{|g|})$, so that they fall into a region where for small $|g|$ the wavefunction $\Psi$ is exponentially small. Therefore these barriers will have no influence on the power expansion.
With
\begin{equation*}
\label{Ansatz}
\Psi=exp(-\Phi),\quad \Phi=\sum_{k=1,j=1}c_{k,j}\;\alpha^k\;x^{2j} \quad \mbox{and}\;e=\sum_{i=1}e_i\,\alpha^i
\end{equation*}
inserted into equation (\ref{Schr2}), a recurrence relation for the $c_{k,j}$ is obtained, which has a unique solution, if one takes the following initial conditions into account:
\begin{itemize}
\item Working up to order $2n$ in $\alpha$ and $x$, we put $c_{k,j}=0$ for $k>2n-1$.
\item The power of $x$ is to be restricted by  $c_{k,j}=0$ for $2j>k+1$.
\end{itemize} 
The results for the coefficients $c_{k,j}$ are listed in {\it table~\ref{TabIII}}. They agree with the literature values \cite{ASt}. The recurrence relation for the $e_i$ can be found in appendix B.
\begin{table}[htp!]
\caption[TabIII]{ The first few coefficients of the strong coupling expansion for the ground state of the $cos(2x)$-potential inside an infinite square well.}
\label{TabIII}
\begin{center}
\begin{tabular}{||c|c||}\hline \hline
\multicolumn{2}{||c||}{\bf Square Well with $Cos(2x)$} \\ \hline
\multicolumn{2}{||c||}{\it Strong Coupling} \\ \hline
$i $ & $ e_i$  \\ \hline \hline
$0 $ & $ 0 $ \\ \hline
$1 $ & $ 1 $ \\ \hline
$2 $ & $ -1/8 $ \\ \hline
$3 $ & $ -1/64 $ \\ \hline
$4 $ & $ -3/512 $ \\ \hline
$5 $ & $ -53/16\,384 $ \\ \hline
$6 $ & $ -297/131\,072 $ \\ \hline
$7 $ & $ -3\,961/2\,097\,152 $ \\ \hline
$8 $ & $ -30\,363/16\,777\,216 $ \\ \hline
$9 $ & $ -2\,095\,501/1\,073\,741\,824 $ \\ \hline
$10 $ & $ -20\,057\,205/8\,589\,934\,592 $ \\ \hline \hline
$11 $ & $ -421\,644\,859/137\,438\,953\,472 $ \\ \hline \hline
$12 $ & $ -4\,828\,704\,237/1\,099\,511\,627\,776 $ \\ \hline \hline
\end{tabular}
\end{center}
\end{table}
This series is not of Borell type; again it has a finite radius of convergence of $1/\sqrt{|g|}\simeq 4$. The agreement of both expansions with numerical results is shown in {\it figure~\ref{FigI}} and in {\it figure~\ref{FigII}} .
\hspace{1cm}
\section{Appendix}
\begin{itemize}
\item {\it A: The Recurrence Relation for the Weak Coupling Coefficients}

\begin{equation*}
\label{A1}
\gamma_{k,i} = \left \{ \begin{array}{l}
1 \quad \mbox{for}\;k=i=0\\
0 \quad \mbox{for}\;k=0\; \mbox{and}\;i \neq 0\; \mbox{or}\;i=0\; \mbox{and}\;k \ne 0\\
\frac{\gamma_{k,i-1}-\gamma_{k-2,i-1}-\gamma_{k+2,i-1}+\sum \limits_{j=2}^{i-1}\gamma_{2,j-1}\gamma_{k,i-j}}{k^2+2k } \quad \mbox{else}\\
\end{array}\right \}
\end{equation*}
\begin{equation*}
\label{A2}
\epsilon_i = \frac{1}{2}\left \{ \begin{array}{ll}
1 & \mbox{for}\;i=0,1\\
\gamma_{2,i-1} \quad & \mbox{else}\\
\end{array}\right \}
\end{equation*}
\item {\it B: The Recurrence Relation for the Strong Coupling Coefficients}

To evalute the strong coupling coefficients $e_i$ up to order $n$ in the expansion parameter $\alpha=1/\sqrt{|g|}$ the following relation may be used:
\begin{equation*}
\label{A3}
\gamma_{k,j} = \left \{ \begin{array}{ll}
1 & \mbox{for}\;k=j=1\\
0 & \mbox{for}\;j<1\; \mbox{or}\;k>2n-1\; \mbox{or}\;2j>k+1\\
\beta_{k,j}+\frac{(-4)^{j-1}}{j(2j)!}\quad & \mbox{for}\;2j=k+1\\
\beta_{k,j} & \mbox{else}\\
\end{array}\right \}
\end{equation*}
where
\begin{equation*}
\label{A4}
\beta_{k,j} =\frac{(j+1)(2j+1)\gamma_{k+1,j+1}-2\sum \limits_{l=2}^{k-1}  \sum \limits_{i=[[j-\frac{k-l+1}{2}]]}^{Min(j,\frac{l+1}{2} )} i(j-i+1)\gamma_{l,i}\,\gamma_{k+1-l,j+1-i}}{4j}
\end{equation*}
and
\begin{equation*}
\label{A5}
e_i =\gamma_{i,1} 
\end{equation*}

\end{itemize}
\newpage
\end{document}